\documentclass[11pt,a4paper]{article}
\usepackage[T1]{fontenc}
\usepackage[utf8]{inputenc}
\usepackage{lmodern}
\usepackage[margin=2.65cm]{geometry}
\usepackage{xcolor}
\usepackage{amsmath,amssymb,amsthm,mathtools,mathrsfs,microtype,cite}
\usepackage[hidelinks]{hyperref}
\hypersetup{pdftitle={Necessary and Sufficient Conditions for Adaptive Stabilizability of Continuous-Time Markov Jump Linear Stochastic Systems}}
\numberwithin{equation}{section}
\allowdisplaybreaks[1]
\newtheorem{theorem}{Theorem}[section]

\newtheorem{lemma}{Lemma}[section]
\theoremstyle{definition}
\newtheorem{definition}{Definition}[section]
\newtheorem{remark}{Remark}[section]
\newcommand{\R}{\mathbb R}
\newcommand{\E}{\mathbb E}
\newcommand{\Prob}{\mathbb P}
\newcommand{\Sset}{\mathcal S}
\newcommand{\ind}{\mathbf 1}
\DeclareMathOperator{\tr}{tr}
\DeclareMathOperator{\rank}{rank}
\DeclareMathOperator{\diag}{diag}

\DeclareMathOperator*{\argmin}{arg\,min}
\newcommand{\doi}[1]{{\urlstyle{same}\href{https://doi.org/#1}{\nolinkurl{#1}}}}
\title{Necessary and Sufficient Conditions for Adaptive Stabilizability\\ of Continuous-Time Markov Jump Linear Stochastic Systems}
\author{Ziwei Guo\thanks{
State Key Laboratory of Mathematical Sciences,
Academy of Mathematics and Systems Science,
Chinese Academy of Sciences, Beijing 100190, China;
School of Mathematical Sciences,
University of Chinese Academy of Sciences,
Beijing 100049, China.
(e-mail: \href{mailto:guoziwei@amss.ac.cn}
{guoziwei@amss.ac.cn}).}}
\date{}
\begin{document}
\maketitle
\vspace{-1.5em}
\begin{abstract}The purpose of this paper is to provide a mathematically rigorous solution to a well-known problem in stochastic adaptive control, i.e., to find the weakest possible structural condition for global stabilizability of continuous-time Markov jump linear stochastic systems with known candidate modes but with an unobservable Markov parameter process.
Under some regularity assumptions, it is shown that the necessary and sufficient condition for global adaptive stabilizability hinges on the solvability of the same coupled Riccati equations as in the non-adaptive case. 
This result highlights a fundamental difference between the continuous-time case considered here and the discrete-time case studied earlier by Xue and Guo (2001), in which the coupled Riccati equations for adaptive and non-adaptive stabilizability are different.

\end{abstract}

\section{Introduction}\label{sec:introduction}
Uncertainty is a fundamental issue in feedback control. While adaptive control exploits observations collected during system operation to cope with unknown dynamics, a more fundamental question than the design of a particular control algorithm is to determine which uncertain systems can be stabilized by feedback under a given uncertainty and information structure~\cite{Guo2020}. 
In practical engineering systems, uncertainty may arise not only from fixed unknown parameters, but also from random changes in system dynamics.
For example, actuator failures in robotic manipulators, generator outages, and transmission-line failures in power systems may cause the system to switch randomly among different operating modes~\cite{SiqueiraTerra2009,LiuMiaoFanHan2017}.
Markov jump linear systems (MJLSs), in which the switching among different linear dynamical modes is governed by a finite-state Markov chain, provide a natural mathematical framework for modeling such phenomena and studying the associated stability and feedback control problems~\cite{CostaFragosoTodorov2013}. 
In many practical situations, however, although the dynamical models associated with the candidate modes and the transition law are known, the currently active mode cannot be directly observed. 
The controller must therefore operate without direct mode information, as in the discrete-time state-feedback speed control of direct-current (DC) motors
subject to abrupt power failures~\cite{VargasCostaDoVal2013}.
This raises a fundamental question: which Markov jump linear systems can be stabilized by adaptive feedback when the operating modes are unobservable?

To investigate the effect of missing mode information on feedback stabilizability, it is natural to first consider the case where the operating mode is fully observable. When both the state and the mode are observed, feedback control, stabilizability, and stochastic stability of Markov jump linear systems have been extensively studied~\cite{Sworder1969,JiChizeck1990,FengEtAl1992}. 
In particular, Fang and Loparo~\cite{FangLoparo2002} established necessary and sufficient conditions for mean-square stabilizability of continuous-time Markov jump linear systems in terms of the existence of positive definite solutions to coupled algebraic Riccati equations. 
Related studies include approaches based on linear matrix inequalities~\cite{AitRamiElGhaoui1996}, as well as stability analysis and coupled Lyapunov equations for systems with additive disturbances~\cite{CostaFragosoTodorov2013,FragosoCosta2005}. 
However, mode-dependent feedback laws generally require knowledge of the current operating mode and therefore cannot be directly implemented when the mode is unobservable and only the state observation history is available.

To address the control difficulties arising from unobservable modes, an important line of research is to estimate the hidden operating modes from the observed state history and construct adaptive feedback accordingly. Caines and Chen~\cite{CainesChen1985} studied adaptive LQG control for systems with finite-state random parameters using nonlinear filtering, while Caines and Nassiri-Toussi~\cite{CainesNassiri1991} further investigated adaptive stabilization and ergodic behavior. Subsequently, Dufour and Bertrand~\cite{DufourBertrand1994} developed a stabilizing control law for hybrid systems under algebraic sufficient conditions, and Caines and Zhang~\cite{CainesZhang1995} constructed certainty-equivalence feedback based on the Wonham filter, establishing time-averaged second-moment bounds under suitable conditions. Dufour and Elliott~\cite{DufourElliott1998} also studied filtering and adaptive control of linear systems with Markov perturbations under a different observation structure. 
These studies provide important approaches to feedback design with hidden modes, but also raise a more fundamental question: what are the weakest conditions under which adaptive stabilization can be achieved, and can these conditions be characterized by a necessary and sufficient criterion for general admissible
causal feedback laws?

In the discrete-time case, the capability of adaptive feedback stabilization has been extensively investigated.
Huang and Guo~\cite{HuangGuo2001} showed that stabilization can be achieved by a gradient-based adaptive controller when the hidden Markov parameters switch sufficiently slowly. Subsequently, Xue, Guo, and Huang~\cite{XueGuoHuang2001} further revealed that adaptive feedback capability depends not only on the rate of parameter variation, but also on information uncertainty and the structural complexity of the system. 
Building on these studies, Xue and Guo~\cite{XueGuo2001} established necessary and sufficient conditions for adaptive stabilizability of finite-dimensional discrete-time jump linear systems under suitable regularity and identifiability assumptions, characterized by the solvability of coupled Riccati-like equations. In particular, they showed that the corresponding criteria for adaptive and non-adaptive stabilizability can differ, indicating that missing mode information may strictly limit feedback stabilization capability. 
Liu and Guo~\cite{LiuGuo2021} further developed a spectral characterization of this problem. These results naturally raise the following question: does the lack of mode information still lead to a loss of feedback stabilization capability in the continuous-time case?

In the continuous-time case, adaptive stabilization of systems with hidden modes has also received considerable attention. 
Tan, Zhang, and Yao~\cite{TanZhangYao2005} investigated adaptive sampled control and its quadratic performance for noise-free systems, while Tan and Zhang~\cite{TanZhang2009} further considered systems with stochastic noise and developed a double-stepped adaptive control scheme.
Bercu, Dufour, and Yin~\cite{BercuDufourYin2009} constructed feedback laws based on posterior mode probabilities and coupled Riccati equations, establishing almost-sure stabilization under
additional positive definiteness conditions.
Liu, Liu, and Dong\cite{LiuLiuDong2010} studied a continuous-time adaptive stabilizability problem, but their criterion relies on a state-dependent Lyapunov integral inequality.
Liu, He, and Li~\cite{LiuHeLi2016} studied adaptive LQG control with hidden Markov modes and stochastic disturbances, proposing a sufficient stabilization condition and a performance bound. However, the distinguishability implication in their Remark~2.4 is invalid under the stated assumptions, so the subsequent estimation-error bound requires additional justification; Appendix~\ref{app:liu2016-counterexample} provides a counterexample.
Despite these advances, it remains to rigorously establish structural
necessary and sufficient conditions for adaptive stabilizability in
continuous time under weaker distinguishability assumptions, and to
determine whether the absence of mode information reduces the
stabilizing capability of feedback.

To address these questions, we consider a class of continuous-time Markov jump linear stochastic systems in which the candidate dynamics and transition law are known, the state is continuously observed, and the operating mode is unobservable. We assume an exogenous irreducible Markov chain and nondegenerate additive Brownian noise independent of the mode process. Our distinguishability assumption is an observability rank condition strictly weaker than the instantaneous full-rank condition imposed by Xue and Guo~\cite{XueGuo2001}; it allows modes to be distinguished through their state trajectories even when the instantaneous drift difference is singular.

Our contributions are twofold. First, under this weaker trajectory-distinguishability condition, we give a constructive proof that positive definite solutions of the standard coupled algebraic Riccati equations imply adaptive stabilizability in the uniform mean-square bounded sense. Second, we establish necessity for general admissible causal feedback laws, thereby proving that adaptive stabilizability, non-adaptive stabilizability, and solvability of these coupled Riccati equations are equivalent. Consequently, in contrast to the discrete-time case, unobservable modes impose no additional restriction on feedback stabilization capability under the stated assumptions. The classical equivalence between the Riccati and spectral conditions~\cite{FangLoparo2002} then yields an equivalent characterization by negativity of the infimum of the spectral abscissa over mode-dependent feedback gains. The resulting necessary and sufficient criterion characterizes the intrinsic stabilizing capability of feedback, rather than the performance of a particular control algorithm.

The remainder of the paper is organized as follows.
Section~\ref{sec:model} introduces the model and basic definitions,
and Section~\ref{sec:results} presents the assumptions and main results.
Section~\ref{sec:proof} provides the proofs of the main results,
and Section~\ref{sec:conclusion} concludes the paper.
Supplementary material is provided in the appendices.

\medskip\noindent\textbf{Notation.}
The symbols $|\cdot|$ and $\|\cdot\|$ denote the Euclidean
vector norm and its induced matrix norm, respectively.
For a square matrix $H$, $\sigma(H)$ denotes its spectrum.
If $H$ is symmetric, $H\succ0$ and $H\succeq0$
mean positive definiteness and positive semidefiniteness,
respectively, while $\lambda_{\min}(H)$ and
$\lambda_{\max}(H)$ denote its smallest and largest
eigenvalues.
The symbol $\ind_E$ denotes the indicator of an event $E$. 
The notation $L^p$ ($p=1,2$) denotes the usual
spaces of $p$-integrable functions or random variables.

\section{Model and Preliminaries}\label{sec:model}

Let $\Sset=\{1,\ldots,N\}$ be the finite set of operating modes.
Let $\theta=(\theta_t)_{t\ge0}$ be a continuous-time Markov chain taking values in $\Sset$, with known generator $\Lambda=(\lambda_{ij})$; $\theta_t$ denotes the active mode at time $t$.
Let $W=(W_t)_{t\ge0}$ be an $r$-dimensional standard Brownian motion (Wiener process), which drives the continuous random disturbances.
We consider the stochastic system
\begin{equation}\label{eq:model}
 dx_t=(A(\theta_t)x_t+B(\theta_t)u_t)\,dt+D\,dW_t,
 \qquad x_0=\xi\in\R^n,
\end{equation}
where $x_t\in\R^n$ and $u_t\in\R^m$ are the system state and control input, respectively, and $\xi$ is the initial state.
The candidate system matrices $A_i\triangleq A(i)\in\R^{n\times n}$ and $B_i\triangleq B(i)\in\R^{n\times m}$ are known, and $D\in\R^{n\times r}$ is a known noise-intensity matrix.
The assumptions on $\theta$ and $W$ are stated in Section~\ref{sec:results}.
We write $\Sigma=DD^\top$ and $\rho=\tr\Sigma$.
For $i\ne j$, $\lambda_{ij}$ is the transition rate from mode $i$ to mode $j$. The row convention is used for the generator:
\begin{equation}\label{eq:generator}
 \Prob(\theta_{t+h}=j\mid\theta_t=i)
 =\delta_{ij}+\lambda_{ij}h+o(h),\qquad h\downarrow0,
\end{equation}
where $\delta_{ij}$ is the Kronecker delta,
$\lambda_{ij}\ge0$ for $j\ne i$, and
$\lambda_{ii}=-\sum_{j\ne i}\lambda_{ij}$.
The state is observed continuously and without measurement error. The term $D\,dW_t$ is a process disturbance, not measurement noise.

For each deterministic initial pair $(\xi,i_0)\in\R^n\times\Sset$, let $\Prob_{\xi,i_0}$ denote the law with $x_0=\xi$ and $\theta_0=i_0$, and write $\E_{\xi,i_0}$ for expectation. Subscripts are omitted when the initial pair and feedback have been fixed. This notation does not give the initial mode to an adaptive controller.

Let $\mathcal G_t$ be the usual augmentation of $\sigma\{W_s,\theta_s:0\le s\le t\}$. For a strong solution, let $\mathcal F_t^x$ and $\mathcal F_t^{x,\theta}$ be the usual augmentations of the filtrations generated by $x_{[0,t]}$ and $(x_{[0,t]},\theta_{[0,t]})$. Thus
\[
 \mathcal F_t^x\subseteq\mathcal F_t^{x,\theta}\subseteq\mathcal G_t.
\]
We use the standard conventions for augmentation and stochastic integration\cite{LiptserShiryayev1977}. A causal path rule is a measurable function of the indicated histories, with no dependence on future observations.

\begin{definition}\label{def:admissible}
A causal control process is an admissible feedback if, for every initial pair, it is progressively measurable with respect to the indicated observation filtration, equation~\eqref{eq:model} has a unique non-explosive strong solution, and
\begin{equation}\label{eq:finite-energy}
 \E_{\xi,i_0}\int_0^T|u_t|^2\,dt<\infty,\qquad T<\infty.
\end{equation}
The control is \emph{adaptive} if $u_t=\mu_t(x_{[0,t]})$ and the filtration is $\mathcal F_t^x$. It is \emph{non-adaptive} if it may have the form $u_t=\nu_t(x_{[0,t]},\theta_{[0,t]})$ and the filtration is $\mathcal F_t^{x,\theta}$. 
\end{definition}

The term ``non-adaptive'' refers to observation of the true mode, as in Xue and Guo\cite{XueGuo2001}. It does not restrict the feedback to be linear or memoryless. An adaptive feedback is a non-adaptive feedback that ignores the mode argument. Neither class observes $W$ directly. The strong solution satisfies, almost surely for every $t\ge0$,
\begin{equation}\label{eq:integral-model}
 x_t=\xi+\int_0^t(A(\theta_s)x_s+B(\theta_s)u_s)\,ds+DW_t.
\end{equation}

\begin{definition}\label{def:stabilizability}
System~\eqref{eq:model} is said to be
\emph{adaptively stabilizable} if there exists
an admissible adaptive feedback such that
\begin{equation}\label{eq:stabilizable}
 \sup_{t\ge0}\E_{\xi,i_0}|x_t|^2<\infty,
 \qquad (\xi,i_0)\in\R^n\times\Sset.
\end{equation}
It is said to be \emph{non-adaptively stabilizable}
if the same condition holds for some admissible
non-adaptive feedback.
\end{definition}

The bound in~\eqref{eq:stabilizable} may depend on
the initial pair.
For the noisy system~\eqref{eq:model}, stabilizability
requires neither
$\lim_{t\to\infty}\E_{\xi,i_0}|x_t|^2=0$
nor
$\sup_{t\ge0}|x_t|<\infty$ almost surely.

\section{Main Results}\label{sec:results}

We make the following assumptions.

\medskip\noindent\textbf{A1.}
The Brownian motion $W$ and the Markov chain $\theta$
are independent, and $\Sigma\succ0$.

\medskip\noindent\textbf{A2.}
The chain $\theta$ is exogenous, irreducible, and
time-homogeneous.

\medskip
Under A2, only finitely many mode jumps occur on each
bounded time interval, almost surely.
The chain has a unique stationary distribution
$\pi=(\pi_1,\ldots,\pi_N)^\top$ satisfying
\begin{equation}\label{eq:stationary}
 \Lambda^\top\pi=0,\qquad
 \sum_{i=1}^N\pi_i=1,\qquad
 \pi_i>0,\quad i\in\Sset.
\end{equation}
Moreover, for any initial pair $(\xi,i_0)\in\R^n\times\Sset$,
\begin{equation}\label{eq:mode-convergence}
 \Prob_{\xi,i_0}(\theta_t=i)\longrightarrow\pi_i,
 \qquad t\to\infty,\quad i\in\Sset.
\end{equation}
See~\cite[Section~20.3]{MeynTweedie2009}.
The initial distribution of $\theta_0$ need not be
the stationary distribution $\pi$.
Set $\bar\lambda=\max_{i\in\Sset}(-\lambda_{ii})$.

\medskip\noindent\textbf{A3.}
There exists $L\in\R^{m\times n}$ such that
\begin{equation}\label{eq:identifiability}
 \rank\mathcal O_{ij}(L)=n,
 \qquad i,j\in\Sset,\quad i\ne j,
\end{equation}
where
\begin{equation}\label{eq:distinguishability-matrix}
 \begin{gathered}
 \mathcal O_{ij}(L)=
 \begin{pmatrix}
  \Delta_{ij}(L)\\
  \Delta_{ij}(L)F_i(L)\\
  \vdots\\
  \Delta_{ij}(L)F_i(L)^{n-1}
 \end{pmatrix},\\
 F_i(L)=A_i-B_iL,\qquad
 \Delta_{ij}(L)=F_i(L)-F_j(L).
 \end{gathered}
\end{equation}

\begin{remark}\label{rem:distinguishability}
Under the common feedback $u=-Lx$ with $L$ satisfying A3,
consider the noise-free system with the true mode fixed
at $i\in\Sset$.
Using the same state $x$ and input $u$, a candidate
mode $j\in\Sset$ predicts the state derivative as
$A_jx+B_ju$.
Taking the prediction error as the output gives
the auxiliary feedback system
\[
\left\{
\begin{aligned}
\dot x&=A_ix+B_iu=F_i(L)x,\\
y_{ij}&=\dot x-(A_jx+B_ju)=\Delta_{ij}(L)x.
\end{aligned}
\right.
\]
A3 is the observability condition for this auxiliary
system when $i\ne j$.
For $x_0=\xi\ne0$, the prediction error therefore
provides a criterion for identifying the true mode:
an identically zero error implies $j=i$, whereas
an error that is not identically zero implies $j\ne i$.
The gain $L$ need not be stabilizing.

The condition in Xue and Guo
\cite[Assumption~A2]{XueGuo2001}
requires the existence of a common gain $\bar L$
such that
\begin{equation}\label{eq:xue-guo-rank}
 \rank\Delta_{ij}(\bar L)=n,
 \qquad i,j\in\Sset,\quad i\ne j.
\end{equation}
This implies~\eqref{eq:identifiability}, since
$\Delta_{ij}(\bar L)$ is the first block of
$\mathcal O_{ij}(\bar L)$.
In contrast, A3 only requires the observability
matrix to have full rank and is therefore more general.
An example showing that A3 is strictly weaker
than~\eqref{eq:xue-guo-rank}
is given in Appendix~\ref{app:observability-example}.
\end{remark}

\medskip
To state the algebraic criteria for stabilizability, consider the coupled algebraic Riccati equations
\begin{equation}\label{eq:riccati}
 A_i^\top P_i+P_iA_i+\sum_{j=1}^N\lambda_{ij}P_j
 -P_iB_iB_i^\top P_i+I_n=0,
 \qquad i\in\Sset.
\end{equation}

To characterize the stabilizability of
system~\eqref{eq:model} through spectral optimization,
define
\begin{equation}\label{eq:optimal-spectral-abscissa}
 s_*=\inf_{G\in(\R^{m\times n})^N}
 s\bigl(\mathscr A(G)\bigr),
\end{equation}
where $G=(G_1,\ldots,G_N)$,
$s(H)=\max\{\operatorname{Re}\mu:\mu\in\sigma(H)\}$
denotes the spectral abscissa of a square matrix $H$, and
$\mathscr A(G)$ is defined by~\cite[Section~3.3]{CostaFragosoTodorov2013}
\begin{equation}\label{eq:spectral-matrix}
 \mathscr A(G)=
 \diag_{i=1}^N\bigl(I_n\otimes(A_i-B_iG_i)
 +(A_i-B_iG_i)\otimes I_n\bigr)
 +\Lambda^\top\otimes I_{n^2}.
\end{equation}
Here $\otimes$ denotes the Kronecker product.
The infimum need not be attained and may equal $-\infty$.

\begin{theorem}\label{thm:main}
Under A1--A3, the following statements are equivalent:
\begin{enumerate}
 \item[\textnormal{(i)}] System~\eqref{eq:model} is adaptively stabilizable.
 \item[\textnormal{(ii)}] System~\eqref{eq:model} is non-adaptively stabilizable.
 \item[\textnormal{(iii)}] The coupled Equations~\eqref{eq:riccati} have a solution set
 $P_i=P_i^\top\succ0$, $i\in\Sset$.
 \item[\textnormal{(iv)}] The infimum of the matrix spectral abscissa defined by ~\eqref{eq:optimal-spectral-abscissa} satisfies
 \begin{equation}\label{eq:spectral-condition}
  s_*<0.
 \end{equation}
\end{enumerate}
\end{theorem}

It can be seen from the proof of Theorem \ref{thm:main} that, when these conditions hold, an admissible adaptive feedback can be chosen so that
\begin{equation}\label{eq:theorem-bound}
 \E_{\xi,i_0}(|x_t|^2+|u_t|^2)
 \le C e^{-\gamma t}|\xi|^2+B,
 \qquad t\ge0,\quad (\xi,i_0)\in\R^n\times\Sset,
\end{equation}
where $C,\gamma>0$ and $B\ge0$ do not depend on the initial pair.

\begin{remark}\label{rem:interpretation}
Theorem~\ref{thm:main} applies to general admissible causal feedback laws,
without restricting them to linear or memoryless forms.
The necessity implication (ii)$\Rightarrow$(iii) requires only A1 and A2;
A3 is used only in the sufficiency proof (iii)$\Rightarrow$(i)
to construct an adaptive stabilizing feedback.
The algebraic equivalence (iii)$\Leftrightarrow$(iv) does not require A3.
It expresses the existence of a positive definite solution to the coupled
Riccati equations as negativity of the infimum of the spectral abscissa
over mode-dependent gain tuples.
Thus the same spectral optimization criterion characterizes both adaptive
and non-adaptive stabilizability under the stated assumptions.
This gives a continuous-time counterpart to the discrete-time
spectral-radius formulation in~\cite{LiuGuo2021}.
\end{remark}

\begin{remark}\label{rem:average}
Under A1--A3, the four conditions in Theorem~\ref{thm:main} remain equivalent if the state bound in Definition~\ref{def:stabilizability} is replaced by
\begin{equation}\label{eq:average-criterion}
 \limsup_{T\to\infty}\frac1T\E_{\xi,i_0}
 \int_0^T(|x_t|^2+|u_t|^2)\,dt<\infty,
 \qquad (\xi,i_0)\in\R^n\times\Sset,
\end{equation}
or by the same condition with only $|x_t|^2$ in the integrand. The criterion~\eqref{eq:average-criterion} is used in Liu et al.\cite[equation~(17)]{LiuLiuDong2010}. The assertion concerns the existence of feedback, not equivalence of the bounds for an arbitrary fixed feedback. Its proof is given in Appendix~\ref{app:average}.
\end{remark}

\section{Proof of the Main Results}\label{sec:proof}
\subsection{Auxiliary lemmas}

The first lemma extends the usual mode-dependent second-moment equation\cite{FragosoCosta2005} to a general history-dependent control, under which the state-mode pair need not be Markov.
Its proof, based on the It\^o product formula
for switching systems~\cite[Section~2.1]{BaranYinZhu2013},
is given in Appendix~\ref{app:moment}.

\begin{lemma}\label{lem:moment}
Under A1--A2, let $u$ be an admissible feedback.
For $i\in\Sset$, define
\begin{equation}\label{eq:moments}
 M_i(t)=\E[x_tx_t^\top\ind_{\{\theta_t=i\}}],
 \quad U_i(t)=\E[u_tx_t^\top\ind_{\{\theta_t=i\}}],
 \quad p_i(t)=\Prob(\theta_t=i).
\end{equation}
Then $M_i(\cdot)$ is locally absolutely continuous,
$U_i(\cdot)$ is locally integrable, and
\begin{equation}\label{eq:moment-equation}
 \begin{aligned}
 \dot M_i(t)
 &=A_iM_i(t)+M_i(t)A_i^\top
   +B_iU_i(t)+U_i(t)^\top B_i^\top\\
 &\quad+\sum_{j=1}^N\lambda_{ji}M_j(t)+p_i(t)\Sigma,
 \end{aligned}
\end{equation}
for almost every $t\ge0$.
The conclusion remains valid without the
nondegeneracy condition $\Sigma\succ0$,
in particular when $D=0$.
\end{lemma}

Combining the Riccati characterization
in~\cite[Corollary~3.1]{FangLoparo2002}
with the LMI characterization
in~\cite[Theorem~2.1]{AitRamiElGhaoui1996}
yields the following equivalence.

\begin{lemma}\label{lem:algebra}
Under A2, equations~\eqref{eq:riccati} have
a positive definite solution if and only if
there exist $X_i=X_i^\top\succ0$ and
$Y_i\in\R^{m\times n}$ satisfying
\begin{equation}\label{eq:matrix-inequality}
 A_iX_i+X_iA_i^\top-B_iY_i-Y_i^\top B_i^\top
 +\sum_{j=1}^N\lambda_{ji}X_j\prec0,\qquad i\in\Sset.
\end{equation}
\end{lemma}

The following lemma gives a conditional moment
formula for linear switching diffusions, based on
the variation-of-constants formula and the It\^o
isometry~\cite[Sections~4.2--4.3]{LiptserShiryayev1977}.
A brief proof for the present setting is given
in Appendix~\ref{app:conditional}.

\begin{lemma}\label{lem:conditional}
Under A1, let $0\le a<b$ and consider
$dz_t=F_{\theta_t,\ell}z_t\,dt+D\,dW_t$,
where $z_a\in L^2(\mathcal G_a;\R^n)$,
$\ell\in\Sset$ is $\mathcal G_a$-measurable,
and the deterministic matrices $F_{i\ell}$ satisfy
$\|F_{i\ell}\|\le M$.
Let $\dot\Psi=F_{\theta_t,\ell}\Psi$, $\Psi(a)=I_n$,
and $\Phi(t,s)=\Psi(t)\Psi(s)^{-1}$.
Then
\begin{equation}\label{eq:lemma-adapted-variation}
 z_t=\Phi(t,a)z_a+\eta_t,\qquad
 \eta_t=\Psi(t)\int_a^t\Psi(s)^{-1}D\,dW_s,
\end{equation}
with $\|\Phi(t,s)\|\le e^{M(t-s)}$.
Moreover, for
$\mathcal H=\mathcal G_a\vee\sigma\{\theta_s:a\le s\le b\}$,
\begin{equation}\label{eq:lemma-conditional-moments}
 \E[\eta_t\mid\mathcal H]=0,\qquad
 \E[\eta_t\eta_t^\top\mid\mathcal H]
 =\int_a^t\Phi(t,s)\Sigma\Phi(t,s)^\top\,ds,
\end{equation}
for $t\in[a,b]$.
\end{lemma}

\subsection{Proof of Theorem~\ref{thm:main}}
\begin{proof}
We first prove (i)$\Rightarrow$(ii)$\Rightarrow$(iii)$\Rightarrow$(i),
and then establish (iii)$\Leftrightarrow$(iv).

\medskip\noindent\textbf{(i)$\Rightarrow$(ii).}
An admissible adaptive rule $u_t=\mu_t(x_{[0,t]})$ is also a non-adaptive rule, by setting
$\nu_t(x_{[0,t]},\theta_{[0,t]})=\mu_t(x_{[0,t]})$.
The trajectories and their second moments are unchanged.

\medskip\noindent\textbf{(ii)$\Rightarrow$(iii).}
Suppose an admissible non-adaptive feedback satisfies~\eqref{eq:stabilizable}. Fix an initial pair $(\xi,i_0)$ and put $M_* =\sup_{t\ge0}\E|x_t|^2<\infty$. Using Lemma~\ref{lem:moment}, define
\begin{equation}\label{eq:time-averages}
 \begin{aligned}
 &Z_i(T)=\frac1T\int_0^T M_i(t)\,dt,\qquad
 Y_i(T)=-\frac1T\int_0^T U_i(t)\,dt,\\
 &\bar p_i(T)=\frac1T\int_0^T p_i(t)\,dt.
 \end{aligned}
\end{equation}
All these matrices are finite for $T<\infty$, and $Z_i(T)\succeq0$. Integration of~\eqref{eq:moment-equation} gives
\begin{align}\label{eq:averaged-identity}
 &A_iZ_i(T)+Z_i(T)A_i^\top-B_iY_i(T)-Y_i(T)^\top B_i^\top
 +\sum_j\lambda_{ji}Z_j(T)\notag\\
 &\hspace{4em}=\frac{M_i(T)-M_i(0)}T-\bar p_i(T)\Sigma.
\end{align}
We have $\|M_i(t)\|\le M_*$ and $\bar p_i(T)\to\pi_i$ by~\eqref{eq:mode-convergence}. Hence
\[
 \max_i\left\|\frac{M_i(T)-M_i(0)}T-\bar p_i(T)\Sigma+\pi_i\Sigma\right\|
 \le\frac{2M_*}{T}+\max_i|\bar p_i(T)-\pi_i|\,\|\Sigma\|
 \longrightarrow0.
\]
Let $q_* =\min_i\pi_i\lambda_{\min}(\Sigma)>0$. Choose one finite $T$ so that the right-hand side of~\eqref{eq:averaged-identity} is at most $-\delta I_n$ for every $i$, with $\delta=q_*/2$.

Set $X_i=Z_i(T)+\varepsilon\pi_iI_n$ and $Y_i=Y_i(T)$. Then $X_i\succ0$, and the change on the left of~\eqref{eq:averaged-identity} is
\begin{equation}\label{eq:matrix-perturbation}
 \varepsilon\pi_i(A_i+A_i^\top)
 +\varepsilon\sum_j\lambda_{ji}\pi_jI_n
 =\varepsilon\pi_i(A_i+A_i^\top).
\end{equation}
Here $\Lambda^\top\pi=0$ was used. For
$C_A=\max_i\pi_i\|A_i+A_i^\top\|$, choose
$\varepsilon=\delta/[2(C_A+1)]$. The perturbation has norm less than $\delta/2$, so~\eqref{eq:matrix-inequality} holds. Lemma~\ref{lem:algebra} gives (iii). No infinite-horizon bound on the control energy or convergence of $Y_i(T)$ is used.

\medskip\noindent\textbf{(iii)$\Rightarrow$(i).}
Let $P_i\succ0$ solve~\eqref{eq:riccati}. We first treat $N\ge2$.

\medskip\noindent\textit{Step 1. Choice of gains and separation of trajectories.}
Put $L_i=B_i^\top P_i$. The Riccati equations imply
\begin{equation}\label{eq:matched}
 \begin{aligned}
 &(A_i-B_iL_i)^\top P_i+P_i(A_i-B_iL_i)\\
 &\quad+\sum_j\lambda_{ij}P_j=-I_n-L_i^\top L_i\preceq-I_n.
 \end{aligned}
\end{equation}
Fix a common gain $L_0$ satisfying A3. We use the gain-perturbation argument of Xue and Guo\cite[Section~4.2, Step~1, equations~(22)--(23)]{XueGuo2001}. Set
\begin{equation}\label{eq:gain-perturbation}
 K_i(\varepsilon)=(1-\varepsilon)L_i+\varepsilon L_0.
\end{equation}
The change in the left-hand side of~\eqref{eq:matched} is
\[
 -\varepsilon\bigl[(L_0-L_i)^\top B_i^\top P_i
                 +P_iB_i(L_0-L_i)\bigr].
\]
Let $C_L=\max_i\|P_iB_i\|\,\|L_0-L_i\|$ and
$\varepsilon_0=(1+4C_L)^{-1}$. Its norm is at most $2\varepsilon C_L<1/2$ whenever $0<\varepsilon\le\varepsilon_0$.

For $i\ne j$ and $\ell\in\Sset$, define the polynomial
\begin{equation}\label{eq:rank-polynomial}
 q_{ij\ell}(\varepsilon)
 =\det\left[
 \mathcal O_{ij}(K_\ell(\varepsilon))^\top
 \mathcal O_{ij}(K_\ell(\varepsilon))
 \right].
\end{equation}
The entries of $\mathcal O_{ij}(K_\ell(\varepsilon))$ are polynomial in $\varepsilon$, and A3 gives $q_{ij\ell}(1)>0$. None of these finitely many polynomials is identically zero. We may therefore choose one $\varepsilon\in(0,\varepsilon_0]$ outside all their real zero sets. Fix this value and write $K_i=K_i(\varepsilon)$. Then
\begin{align}
 &(A_i-B_iK_i)^\top P_i+P_i(A_i-B_iK_i)
   +\sum_j\lambda_{ij}P_j\preceq-\tfrac12I_n,
   \label{eq:stable-gains}\\
 &\rank\mathcal O_{ij}(K_\ell)=n,
   \qquad i\ne j,\quad \ell\in\Sset.
   \label{eq:distinguishable-gains}
\end{align}
These gains are fixed before the estimation period is chosen. Put
\begin{equation}\label{eq:fixed-constants}
 \begin{aligned}
 &F_{i\ell}=A_i-B_iK_\ell,\qquad
 M=\max\{1,\max_{i,\ell}\|F_{i\ell}\|\},\\
 &a_0=e^{2M},\qquad b_0=\rho a_0.
 \end{aligned}
\end{equation}

For $h>0$, let
\begin{equation}\label{eq:separation-gramian}
 Q_{ij\ell}(h)=\int_0^h
 (e^{F_{i\ell}s}-e^{F_{j\ell}s})^\top
 (e^{F_{i\ell}s}-e^{F_{j\ell}s})\,ds.
\end{equation}
We show that $Q_{ij\ell}(h)\succ0$ for $i\ne j$. If
$z^\top Q_{ij\ell}(h)z=0$, continuity and nonnegativity of the integrand imply
$e^{F_{i\ell}s}z=e^{F_{j\ell}s}z$ for all $s\in[0,h]$. Differentiating the two equal trajectories gives
$(F_{i\ell}-F_{j\ell})e^{F_{i\ell}s}z=0$.
Successive derivatives at $s=0$ yield
\[
 (F_{i\ell}-F_{j\ell})F_{i\ell}^qz=0,
 \qquad q=0,\ldots,n-1.
\]
Thus $\mathcal O_{ij}(K_\ell)z=0$, and~\eqref{eq:distinguishable-gains} gives $z=0$.
Conversely, if the displayed identities hold, the Cayley--Hamilton theorem extends them to every $q\ge0$. The exponential series then gives
$(F_{i\ell}-F_{j\ell})e^{F_{i\ell}s}z=0$ for every $s$. Hence $e^{F_{i\ell}s}z$ solves both linear equations with the same initial value, and the trajectories agree. 

Define, for $h>0$,
\begin{equation}\label{eq:gramian-lower}
 g(h)=\min_{i\ne j,\ell}\lambda_{\min}(Q_{ij\ell}(h))>0.
\end{equation}
For all indicated indices,
\begin{equation}\label{eq:trajectory-separation}
 \int_0^h|(e^{F_{i\ell}s}-e^{F_{j\ell}s})z|^2\,ds
 \ge g(h)|z|^2,
 \qquad i\ne j,\quad\ell\in\Sset.
\end{equation}
No lower bound on $g(h)$ independent of $h$ is assumed.

\medskip\noindent\textit{Step 2. Adaptive feedback.}
For now, let $0<h\le1$, $t_k=kh$, and $x_k=x_{t_k}$, $k\ge0$. Set $\widehat\theta_0=1$ and define
\begin{align}
 u_t&=-K_{\widehat\theta_k}x_t,
       \qquad t\in[t_k,t_{k+1}),
       \label{eq:control}\\
 \widehat\theta_{k+1}
 &=\min\argmin_{i\in\Sset}
   \int_0^h\left|x_{t_k+s}-e^{F_{i,\widehat\theta_k}s}x_k\right|^2\,ds.
       \label{eq:estimator}
\end{align}
The smallest minimizing index is chosen in a tie. The estimate is computed at $t_{k+1}$ from the state interval just observed. Only the gain index is held fixed on an interval; the feedback uses the current state $x_t$.

On each interval, the drift has a fixed gain index and is adapted, uniformly Lipschitz in $x$, and of linear growth. Subtracting the additive Brownian term reduces the equation to a random integral equation with a uniformly Lipschitz drift. Pathwise Picard iteration gives a unique continuous solution, and the iterates are adapted, so the solution is strong; compare the standard Picard construction in\cite[Section~4.4, proof of Theorem~4.6]{LiptserShiryayev1977}. For a fixed previous index, each integral in~\eqref{eq:estimator} is a continuous function of the observed continuous path in the uniform norm. The finite minimization with the stated tie rule is Borel measurable. The solution and estimate can therefore be constructed successively. The resulting drift matrix
\begin{equation}\label{eq:actual-drift}
 F(t)=F_{\theta_t,\widehat\theta_{\lfloor t/h\rfloor}}
\end{equation}
satisfies $\|F(t)\|\le M$. The integral equation, Gronwall's inequality, and Doob's inequality give
\begin{align}\label{eq:closed-local}
 \sup_{t\le T}|x_t|
 &\le e^{MT}\left(|\xi|+\sup_{t\le T}|DW_t|\right),\notag\\
 \E\sup_{t\le T}|x_t|^2
 &\le2e^{2MT}(|\xi|^2+4\rho T)<\infty,
 \qquad T<\infty.
\end{align}
Thus the solution is non-explosive. With $K_* =\max_i\|K_i\|$, we have $|u_t|\le K_*|x_t|$ and
\[
 \E\int_0^T|u_t|^2\,dt
 \le K_*^2T\E\sup_{t\le T}|x_t|^2<\infty.
\]
The control is right-continuous and adapted to $\mathcal F_t^x$, hence progressively measurable. Equations~\eqref{eq:control}--\eqref{eq:estimator} define one admissible adaptive rule for every initial pair.

\medskip\noindent\textit{Step 3. Second moments on the mismatch event.}
Set
\begin{equation}\label{eq:error-definitions}
 f(t)=\E|x_t|^2,\qquad
 E_k=\{\widehat\theta_k\ne\theta_{t_k}\},\qquad
 e_k=\E[|x_k|^2\ind_{E_k}].
\end{equation}
For $t\in[t_k,t_{k+1}]$, let $J_k(t)$ be the event of at least one mode jump in $(t_k,t]$. A jump at a deterministic grid time has probability zero; we exclude this countable union of null events throughout the argument. Put $\ell=\widehat\theta_k$ and
\[
 \mathcal H_k=\mathcal G_{t_k}\vee
 \sigma\{\theta_s:t_k\le s\le t_{k+1}\}.
\]
Lemma~\ref{lem:conditional} gives
\begin{equation}\label{eq:conditional-variation}
 x_t=\Phi_k(t,t_k)x_k+\eta_{k,t},
\end{equation}
where, conditionally on $\mathcal H_k$, the noise term has mean zero and
\begin{align}\label{eq:noise-bound}
 \E[|\eta_{k,t}|^2\mid\mathcal H_k]
 &=\int_{t_k}^t\tr\bigl(\Phi_k(t,s)\Sigma\Phi_k(t,s)^\top\bigr)\,ds\notag\\
 &\le b_0(t-t_k).
\end{align}
Since $x_k$ and $\Phi_k(t,t_k)$ are $\mathcal H_k$-measurable, the cross term vanishes after conditioning. Consequently,
\begin{equation}\label{eq:conditional-state}
 \E[|x_t|^2\mid\mathcal H_k]
 \le a_0|x_k|^2+b_0(t-t_k).
\end{equation}

\textit{Case 1. Intervals with a mode jump.}
The holding-time law gives
\begin{equation}\label{eq:jump-probability}
 \Prob(J_k(t)\mid\mathcal G_{t_k})
 =1-e^{\lambda_{\theta_{t_k}\theta_{t_k}}(t-t_k)}
 \le\bar\lambda(t-t_k).
\end{equation}
The event $J_k(t)$ belongs to $\mathcal H_k$. Applying the tower property to~\eqref{eq:conditional-state}, and then~\eqref{eq:jump-probability}, yields
\begin{align}\label{eq:jump-moment}
 \E[|x_t|^2\ind_{J_k(t)}\mid\mathcal G_{t_k}]
 &\le\bigl(a_0|x_k|^2+b_0(t-t_k)\bigr)
       \Prob(J_k(t)\mid\mathcal G_{t_k})\notag\\
 &\le a_0\bar\lambda h|x_k|^2+b_0\bar\lambda h^2.
\end{align}
In particular,
\begin{equation}\label{eq:endpoint-jump}
 \E[|x_{k+1}|^2\ind_{J_k(t_{k+1})}]
 \le a_0\bar\lambda h f(t_k)+b_0\bar\lambda h^2.
\end{equation}
This estimate does not use independence of the state and the jump event.

\textit{Case 2. An incorrect estimate without a mode jump.}
Write $J_k=J_k(t_{k+1})$ and $i=\theta_{t_k}$. On $J_k^c$,
\begin{equation}\label{eq:no-jump}
 x_{t_k+s}=e^{F_{i\ell}s}x_k+\eta_{k,t_k+s},
 \qquad 0\le s\le h.
\end{equation}
Suppose $E_{k+1}$ also occurs and the estimator chooses $j\ne i$.
Let $r_q(s)=x_{t_k+s}-e^{F_{q\ell}s}x_k$.
The minimizing property gives $\|r_j(\cdot)\|_{L^2(0,h)}\le\|r_i(\cdot)\|_{L^2(0,h)}$.
Since $r_j(s)-r_i(s)=(e^{F_{i\ell}s}-e^{F_{j\ell}s})x_k$, the triangle inequality and~\eqref{eq:trajectory-separation} imply
\begin{align}\label{eq:wrong-state}
 g(h)|x_k|^2
 &\le\int_0^h|(e^{F_{i\ell}s}-e^{F_{j\ell}s})x_k|^2\,ds\notag\\
 &\le4\int_0^h|\eta_{k,t_k+s}|^2\,ds
 \qquad\text{on }E_{k+1}\cap J_k^c.
\end{align}
This is the integral version of the residual comparison in Xue and Guo\cite[Section~4.2, Steps~2--3, equations~(26)--(30)]{XueGuo2001}. At the interval endpoint,~\eqref{eq:no-jump} and~\eqref{eq:wrong-state} give
\begin{equation}\label{eq:wrong-endpoint}
 |x_{k+1}|^2
 \le2a_0|x_k|^2+2|\eta_{k,t_{k+1}}|^2
 \le\frac{8a_0}{g(h)}\int_0^h|\eta_{k,t_k+s}|^2\,ds
      +2|\eta_{k,t_{k+1}}|^2
\end{equation}
on $E_{k+1}\cap J_k^c$.

Because $J_k^c\in\mathcal H_k$, conditional expectation and Tonelli's theorem give
\begin{align}\label{eq:integrated-noise}
 \E\left[\ind_{J_k^c}\int_0^h|\eta_{k,t_k+s}|^2\,ds\right]
 &=\int_0^h\E\left[\ind_{J_k^c}
     \E[|\eta_{k,t_k+s}|^2\mid\mathcal H_k]\right]ds
 \le\frac{b_0h^2}{2},\notag\\
 \E[|\eta_{k,t_{k+1}}|^2\ind_{J_k^c}]&\le b_0h.
\end{align}
Multiplying~\eqref{eq:wrong-endpoint} by the error-event indicator, dropping that indicator from the nonnegative noise terms, and using~\eqref{eq:integrated-noise}, we obtain
\begin{equation}\label{eq:no-jump-error}
 \E[|x_{k+1}|^2\ind_{E_{k+1}\cap J_k^c}]
 \le\frac{4a_0b_0h^2}{g(h)}+2b_0h.
\end{equation}
Let
\begin{equation}\label{eq:grid-constants}
 a_1=a_0\bar\lambda,\qquad
 R_h=b_0\bar\lambda h^2+\frac{4a_0b_0h^2}{g(h)}+2b_0h.
\end{equation}
Combining~\eqref{eq:endpoint-jump} and~\eqref{eq:no-jump-error} proves
\begin{equation}\label{eq:grid-error}
 e_{k+1}\le a_1h f(t_k)+R_h,\qquad k\ge0.
\end{equation}
The constant $R_h$ is finite for every fixed $h>0$.

\medskip\textit{Extension to inter-update times.}
We next extend the grid-point error estimate
\eqref{eq:grid-error} to the intervals between
successive updates.
For $t\in[t_k,t_{k+1})$ and $k\ge1$,
\begin{equation}\label{eq:error-inclusion}
 \{\widehat\theta_k\ne\theta_t\}\subset E_k\cup J_k(t).
\end{equation}
Since $E_k\in\mathcal G_{t_k}$, equation~\eqref{eq:conditional-state} implies
\[
 \E[|x_t|^2\ind_{E_k}]\le a_0e_k+b_0h.
\]
Define $e(t)=\E[|x_t|^2\ind_{\{\widehat\theta_k\ne\theta_t\}}]$ on this interval.
Equations~\eqref{eq:jump-moment} and~\eqref{eq:grid-error} yield
\begin{align}\label{eq:between-grid-error}
 e(t)&\le a_0e_k+b_0h+a_1h f(t_k)+b_0\bar\lambda h^2\notag\\
 &\le a_0a_1h f(t_{k-1})+a_1h f(t_k)
       +a_0R_h+b_0h+b_0\bar\lambda h^2.
\end{align}
By Lemma~\ref{lem:moment}, $f(t)=\sum_i\tr M_i(t)$ is locally absolutely continuous. Summing the traces of~\eqref{eq:moment-equation} cancels the generator terms and gives
\begin{equation}\label{eq:f-lower}
 f'(t)=\E[x_t^\top(F(t)+F(t)^\top)x_t]+\rho\ge-2Mf(t).
\end{equation}
Thus $f(s)\le e^{2M(t-s)}f(t)$ for $s\le t$. Since $h\le1$,
$f(t_k)\le a_0f(t)$ and $f(t_{k-1})\le a_0^2f(t)$.
Set
\begin{equation}\label{eq:error-constants}
 c_1=a_1(a_0^3+a_0),\qquad
 C_h=a_0R_h+b_0h+b_0\bar\lambda h^2.
\end{equation}
Substitution in~\eqref{eq:between-grid-error} gives
\begin{equation}\label{eq:key-estimate}
 \E\left[|x_t|^2
 \ind_{\{\widehat\theta_{\lfloor t/h\rfloor}\ne\theta_t\}}\right]
 \le c_1h\E|x_t|^2+C_h,\qquad t\ge h.
\end{equation}
Here $c_1$ is independent of $h\in(0,1]$, time, and the initial pair. The constant $C_h$ is independent of time and the initial pair, but may grow as $h$ decreases. Only its finiteness at the chosen period will be used.

\medskip\noindent\textit{Step 4. Uniform second-moment bounds.}
Let
\[
 V(x,i)=x^\top P_ix,\qquad v(t)=\E V(x_t,\theta_t),\qquad
 p_- =\min_i\lambda_{\min}(P_i),\quad p_+ =\max_i\lambda_{\max}(P_i).
\]
Then
\begin{equation}\label{eq:lyapunov-bounds}
 p_- f(t)\le v(t)\le p_+f(t).
\end{equation}
Since $F_{i\ell}=F_{ii}+B_i(K_i-K_\ell)$, put
$\kappa=2\max_{i,\ell}\|P_iB_i\|\,\|K_i-K_\ell\|$.
Equation~\eqref{eq:stable-gains} gives
\begin{align}\label{eq:lyapunov-expansion}
 &F_{i\ell}^\top P_i+P_iF_{i\ell}+\sum_j\lambda_{ij}P_j\notag\\
 &=F_{ii}^\top P_i+P_iF_{ii}+\sum_j\lambda_{ij}P_j +(K_i-K_\ell)^\top B_i^\top P_i+P_iB_i(K_i-K_\ell)\notag\\
 &\preceq-\tfrac12I_n+\kappa\ind_{\{\ell\ne i\}}I_n.
\end{align}
Let $c_D=\max_i\tr(P_i\Sigma)$. By Lemma~\ref{lem:moment},
$v(t)=\sum_i\tr(P_iM_i(t))$ is locally absolutely continuous. Differentiating this expression and using~\eqref{eq:lyapunov-expansion} yields
\begin{align}\label{eq:v-drift}
 v'(t)
 &=\E\left[x_t^\top\left(F(t)^\top P_{\theta_t}+P_{\theta_t}F(t)
                  +\sum_j\lambda_{\theta_tj}P_j\right)x_t\right]
   +\sum_i p_i(t)\tr(P_i\Sigma)\notag\\
 &\le-\tfrac12 f(t)+\kappa e(t)+c_D.
\end{align}
The finite-horizon integrability follows from~\eqref{eq:closed-local}. The true mode in $V$ is used only in the analysis. Since $V$ does not depend on the estimated index, estimator updates cause no jump in~$V$.

For $t\ge h$, equation~\eqref{eq:key-estimate} implies
\[
 v'(t)\le-(1/2-\kappa c_1h)f(t)+c_D+\kappa C_h.
\]
If $\kappa c_1>0$, choose one fixed period satisfying
\begin{equation}\label{eq:h-choice}
 0<h\le\min\{1,(4\kappa c_1)^{-1}\}.
\end{equation}
If $\kappa c_1=0$, any $0<h\le1$ suffices. Put
$\gamma=(4p_+)^{-1}$ and $\beta_h=c_D+\kappa C_h$. Then
\begin{equation}\label{eq:final-differential}
 v'(t)\le-\tfrac14 f(t)+\beta_h
         \le-\gamma v(t)+\beta_h,
 \qquad t\ge h.
\end{equation}
Integration gives
\begin{equation}\label{eq:final-v}
 v(t)\le e^{-\gamma(t-h)}v(h)
       +\frac{\beta_h}{\gamma}(1-e^{-\gamma(t-h)}),\qquad t\ge h.
\end{equation}
For the initial interval,~\eqref{eq:conditional-state} gives
\begin{equation}\label{eq:initial-estimate}
 f(t)\le a_0|\xi|^2+b_0h,\quad 0\le t\le h,
 \qquad v(h)\le p_+(a_0|\xi|^2+b_0h).
\end{equation}
Consequently, for $t\ge h$,
\[
 f(t)\le\frac{p_+}{p_-}a_0e^{\gamma h}e^{-\gamma t}|\xi|^2
       +\frac{p_+}{p_-}b_0h+\frac{\beta_h}{\gamma p_-}.
\]
Define
\begin{equation}\label{eq:bound-constants}
 c=\frac{p_+}{p_-}a_0e^\gamma,\qquad
 b_h=\frac{p_+}{p_-}b_0h+\frac{\beta_h}{\gamma p_-}.
\end{equation}
Since $h\le1$, these constants also cover~\eqref{eq:initial-estimate}. Thus
\begin{equation}\label{eq:state-bound}
 \E_{\xi,i_0}|x_t|^2\le c e^{-\gamma t}|\xi|^2+b_h,
 \qquad t\ge0.
\end{equation}
The control satisfies $|u_t|\le K_*|x_t|$, so
\begin{equation}\label{eq:state-control-bound}
 \E_{\xi,i_0}(|x_t|^2+|u_t|^2)
 \le(1+K_*^2)\bigl(c e^{-\gamma t}|\xi|^2+b_h\bigr).
\end{equation}
The gains, period, and constants are independent of the initial pair. Taking
$C=(1+K_*^2)c$ and $B=(1+K_*^2)b_h$ proves~\eqref{eq:theorem-bound} for $N\ge2$.

If $N=1$, A3 is vacuous. Use $u_t=-B_1^\top P_1x_t$. This bounded-gain feedback is admissible by the same finite-horizon estimate. Equation~\eqref{eq:matched} gives, with $p_\pm$ the extreme eigenvalues of~$P_1$,
\begin{equation}\label{eq:single-mode}
 \E_{\xi,1}|x_t|^2
 \le\frac{p_+}{p_-}e^{-t/p_+}|\xi|^2
 +\frac{p_+\tr(P_1\Sigma)}{p_-}(1-e^{-t/p_+}).
\end{equation}
The corresponding bound for state and control follows by multiplying by
$1+\|B_1^\top P_1\|^2$. This proves (iii)$\Rightarrow$(i), establishes the equivalence of (i)--(iii), and gives~\eqref{eq:theorem-bound}.

\medskip\noindent\textbf{(iii)$\Leftrightarrow$(iv).}
By~\cite[Corollary~3.1]{FangLoparo2002}, with state and input weights $I_n$ and $I_m$, condition (iii) is equivalent
to the existence of a gain tuple $G$ for which the auxiliary homogeneous
system with drift $A_i-B_iG_i$ in mode $i$ is mean-square stable.
By~\cite[Theorem~3.6]{FangLoparo2002}, this is equivalent to
$\mathscr A(G)$ being Hurwitz for some $G$:
under our row-generator convention, the matrix used in that theorem is
$\mathscr A(G)^\top$, which has the same spectrum.
Consequently, (iii) holds if and only if $s(\mathscr A(G))<0$
for some $G$, which is equivalent to $s_*<0$ by the definition of the infimum.
This equivalence does not require the infimum to be attained
and remains valid when $s_*=-\infty$.
Together with the preceding implications, this proves all four conditions equivalent.
\end{proof}

\section{Conclusion}\label{sec:conclusion}

We have established a structural necessary and sufficient condition for adaptive stabilizability of continuous-time Markov jump linear stochastic systems with unobservable modes. Under the stated regularity assumptions and an observability-based distinguishability condition strictly weaker than the instantaneous full-rank condition of Xue and Guo~\cite{XueGuo2001}, we have constructed an admissible adaptive feedback with uniformly bounded state and control second moments. Together with a necessity argument valid for general admissible causal feedback, this proves that adaptive and non-adaptive stabilizability are both characterized by the existence of positive definite solutions to the standard coupled algebraic Riccati equations, or equivalently by $s_*<0$. Thus, in the continuous-time setting considered here, missing mode information does not reduce the stabilizing capability of feedback, in contrast to the discrete-time case.

\appendix
\makeatletter
\renewcommand{\@seccntformat}[1]{Appendix~\csname the#1\endcsname\quad}
\makeatother

\section{A Counterexample to Remark~2.4 of Liu et al. (2016)}
\label{app:liu2016-counterexample}

We use the notation of~\cite{LiuHeLi2016}.
Remark~2.4 of that paper asserts that Assumption~2.1
ensures the nonsingularity of
\begin{equation}\label{eq:liu2016-separation}
 D_{ij\ell}(h)
 =e^{(A_i-B_iL_\ell)h}-e^{(A_j-B_jL_\ell)h},
 \qquad i\ne j,
\end{equation}
for the selected feedback gains $L_\ell$ and all
sufficiently small $h>0$.
The following example satisfies Assumptions~2.1--2.2
and the positive definite solution requirement in
equation~(8) of that paper, but contradicts this assertion.

Consider a scalar system with two modes and
\begin{equation}\label{eq:liu2016-data}
 \begin{gathered}
  A_1=0,\qquad A_2=1,\qquad B_1=1,\qquad B_2=2,\\
  \Lambda=\begin{pmatrix}-1&1\\1&-1\end{pmatrix},
  \qquad C_1=C_2=1,
 \end{gathered}
\end{equation}
where $C_i$ denotes the diffusion coefficient.
Let the Brownian motion and the Markov chain be independent.
Taking $\bar L=0$ gives
\[
 (A_1-A_2)-(B_1-B_2)\bar L=-1\ne0,
\]
and the reverse ordered pair gives $1\ne0$.
Thus Assumption~2.1 holds.

Choose
\begin{equation}\label{eq:liu2016-gains-solution}
 L_1=1,\qquad L_2=2,\qquad
 M_1=\frac25,\qquad M_2=\frac15.
\end{equation}
For these gains, equation~(8) of~\cite{LiuHeLi2016}
reduces to
\[
 -3M_1+M_2=-1,\qquad M_1-7M_2=-1.
\]
The coefficient matrix has determinant $20$, so the
stated $M_1,M_2$ form its unique positive solution.

Furthermore, take
\begin{equation}\label{eq:liu2016-weights}
 R_1=\frac25,\qquad R_2=\frac15,\qquad
 Q_1=\frac35,\qquad Q_2=\frac15.
\end{equation}
Since $Q_i>0$, the scalar observability requirement
in Assumption~2.2 holds, and its independence requirement
holds by construction.
Direct substitution also verifies
$L_i=R_i^{-1}B_i^\top M_i$, $i=1,2$, and the coupled
Riccati equations~(25) of~\cite{LiuHeLi2016}.
Thus the example is compatible with the LQ gain
construction as well.

However, under $L_1=1$,
\[
 A_1-B_1L_1=A_2-B_2L_1=-1,
\]
and hence
\begin{equation}\label{eq:liu2016-zero-separation}
 D_{121}(h)=e^{-h}-e^{-h}=0,
 \qquad \det D_{121}(h)=0,
 \qquad h>0.
\end{equation}
Therefore, no $h_0>0$ makes the nonsingularity assertion
in Remark~2.4 valid for these gains.
The existence of a common gain satisfying Assumption~2.1
does not imply the same distinguishability property
for the gains actually selected.

Consequently, the threshold $h_0$ used in
Theorems~3.1--3.2 of~\cite{LiuHeLi2016} cannot be obtained
from Remark~2.4 under the stated assumptions.
The positive separation needed for the estimation-error
bound in Lemma~3.3 is likewise not guaranteed.
That bound is used in estimates~(21)--(22) of the
stabilization proof, on which the subsequent performance
argument also relies.
These steps therefore require additional justification
or a modified gain construction.

\section{An Example of the Observability Condition}\label{app:observability-example}
Consider the two-mode system with
\[
 A_1=A_2=
 \begin{pmatrix}0&1\\0&0\end{pmatrix},
 \qquad
 B_1=\begin{pmatrix}0\\1\end{pmatrix},
 \qquad
 B_2=\begin{pmatrix}0\\-1\end{pmatrix}.
\]
For every $L$, $\Delta_{12}(L)$ has rank at most one,
so~\eqref{eq:xue-guo-rank} fails.
However, $L_0=(1,0)$ gives
\[
 \Delta_{12}(L_0)=
 \begin{pmatrix}0&0\\-2&0\end{pmatrix},
 \qquad
 \Delta_{12}(L_0)F_1(L_0)=
 \begin{pmatrix}0&0\\0&-2\end{pmatrix}.
\]
Hence $\rank\mathcal O_{12}(L_0)=2$, and similarly
$\rank\mathcal O_{21}(L_0)=2$.
Thus A3 holds, showing that it is strictly weaker
than the condition in Xue and Guo\cite{XueGuo2001}.
In particular, at $x=(0,1)^\top$, the two modes
have identical instantaneous drifts, but their
subsequent closed-loop trajectories differ.
Thus, mode distinguishability can be achieved
through state evolution even when the drift
difference matrix is rank deficient.

\section{Proof of Remark~\ref{rem:average}}\label{app:average}
\begin{proof}
Suppose an admissible non-adaptive feedback satisfies
$\limsup_{T\to\infty}T^{-1}\int_0^T f(t)\,dt<\infty$ for each initial pair. Fix one such pair $(\xi,i_0)$. There is $B_0>0$ such that $\int_0^T f(t)\,dt\le B_0T$ for all sufficiently large $T$. The function $f$ is continuous by Lemma~\ref{lem:moment}. For every sufficiently large integer $q$, the inequality
\[
 \frac1q\int_q^{2q}f(t)\,dt\le2B_0
\]
gives some $T_q\in[q,2q]$ with $f(T_q)\le2B_0$. Therefore
$\|M_i(T_q)\|/T_q\to0$, while $\bar p_i(T_q)\to\pi_i$.
The right-hand side of~\eqref{eq:averaged-identity} tends to $-\pi_i\Sigma$ along this sequence. The perturbation~\eqref{eq:matrix-perturbation}, at one sufficiently large $T_q$, gives~\eqref{eq:matrix-inequality}. Lemma~\ref{lem:algebra} proves Riccati solvability.

Conversely, integrating~\eqref{eq:state-control-bound}, dividing by $T$, and letting $T\to\infty$ gives~\eqref{eq:average-criterion}. Equation~\eqref{eq:single-mode} gives the same conclusion for $N=1$. A bound for state and control implies the state-only bound, and adaptive feedback is a subclass of non-adaptive feedback. These observations prove the equivalence of (i)--(iii) for the time-averaged criteria. The algebraic equivalence (iii)$\Leftrightarrow$(iv) is unchanged, so all four conditions remain equivalent.
\end{proof}

\section{Proof of Lemma~\ref{lem:moment}}
\label{app:moment}

\begin{proof}
Fix $T>0$ and set
$a_*=\max_i\|A_i\|$ and $b_*=\max_i\|B_i\|$.
By the integral form of~\eqref{eq:model},
Gronwall's, Cauchy--Schwarz, and Doob's
inequalities, we obtain
\begin{equation}\label{eq:local-moment}
 \E\sup_{0\le t\le T}|x_t|^2
 \le 3e^{2a_*T}\left(
 |\xi|^2+b_*^2T\E\int_0^T|u_t|^2\,dt
 +4\rho T\right)<\infty.
\end{equation}
Hence
\begin{equation}\label{eq:cross-integrability}
 \begin{aligned}
 \E\int_0^T|x_t|^2\,dt
 &\le T\E\sup_{0\le t\le T}|x_t|^2<\infty,\\
 \E\int_0^T|u_t||x_t|\,dt
 &\le
 \left(\E\int_0^T|u_t|^2\,dt\right)^{1/2}
 \left(\E\int_0^T|x_t|^2\,dt\right)^{1/2}<\infty.
 \end{aligned}
\end{equation}
In particular, $M_i$ is bounded on $[0,T]$.
By Tonelli's theorem, $U_i(t)$ is defined for almost
every $t$ and
\[
 \int_0^T\|U_i(t)\|\,dt
 \le\E\int_0^T|u_t||x_t|\,dt<\infty.
\]

Let $I_i(t)=\ind_{\{\theta_t=i\}}$, and write $I_i(t-)$
for its left limit.
For $a,b\in\Sset$, $a\ne b$, let $C_{ab}(t)$ count
the transitions from $a$ to $b$ in $(0,t]$.
Its predictable compensator is
$\int_0^t\lambda_{ab}I_a(s-)\,ds$; see
\cite[Section~2.5]{CostaFragosoTodorov2013}.
Thus
\[
 \widetilde C_{ab}(t)
 =C_{ab}(t)-\int_0^t\lambda_{ab}I_a(s-)\,ds
\]
is a martingale in the augmented natural filtration
of $\theta$, and also in $(\mathcal G_t)$ by independence.
Define
\[
 N_i(t)=\sum_{j\ne i}
 \bigl(\widetilde C_{ji}(t)-\widetilde C_{ij}(t)\bigr).
\]
Since $I_i(t)-I_i(0)=\sum_{j\ne i}(C_{ji}(t)-C_{ij}(t))$,
we obtain
\begin{equation}\label{eq:indicator}
 dI_i(t)=\sum_j\lambda_{ji}I_j(t-)\,dt+dN_i(t).
\end{equation}

For $X_t=x_tx_t^\top$, It\^o's formula
(see also~\cite[Section~2.1]{BaranYinZhu2013}) yields
\begin{equation}\label{eq:xx-ito}
 \begin{aligned}
 dX_t={}&\bigl(
 A(\theta_t)X_t+X_tA(\theta_t)^\top
 +B(\theta_t)u_tx_t^\top\\
 &\quad+x_tu_t^\top B(\theta_t)^\top+\Sigma
 \bigr)\,dt
 +D\,dW_t\,x_t^\top+x_t\,dW_t^\top D^\top.
 \end{aligned}
\end{equation}
Since $X$ is continuous and $I_i$ has finite variation
on $[0,T]$, their quadratic covariation vanishes.
Applying the product formula to $X_tI_i(t)$,
substituting~\eqref{eq:indicator} and~\eqref{eq:xx-ito},
and integrating from $0$ to $t$ gives
\begin{equation}\label{eq:product-integral}
 \begin{aligned}
 X_tI_i(t)={}&X_0I_i(0)
 +\int_0^t\Bigl[
 (A_iX_s+X_sA_i^\top+B_iu_sx_s^\top\\
 &\qquad+x_su_s^\top B_i^\top+\Sigma)I_i(s)
 +\sum_j\lambda_{ji}X_sI_j(s)\Bigr]\,ds\\
 &+\int_0^t I_i(s-)D\,dW_s\,x_s^\top
 +\int_0^t I_i(s-)x_s\,dW_s^\top D^\top\\
 &+\int_0^tX_s\,dN_i(s),\qquad 0\le t\le T.
 \end{aligned}
\end{equation}
Here $I_j(s-)$ has been replaced by $I_j(s)$ only
in the time integrals, since they differ only at
jump times.

The Brownian integrals in~\eqref{eq:product-integral}
have predictable integrands. For $1\le p,q\le n$,
\[
 \E\int_0^T\sum_{\alpha=1}^r
 |I_i(s-)D_{p\alpha}(x_s)_q|^2\,ds
 \le\rho\E\int_0^T|x_s|^2\,ds<\infty.
\]
Thus the Brownian integral
$\int_0^t I_i(s-)D\,dW_s\,x_s^\top$ and its transpose
are square-integrable martingales starting at zero
\cite[Section~4.2]{LiptserShiryayev1977}. Therefore,
\[
 \E\int_0^t I_i(s-)D\,dW_s\,x_s^\top=0,
 \qquad
 \E\int_0^t I_i(s-)x_s\,dW_s^\top D^\top=0.
\]

For the jump integral, $X$ is predictable and
\[
 \E\int_0^T\|X_s\|\lambda_{ab}I_a(s-)\,ds
 \le\lambda_{ab}\E\int_0^T|x_s|^2\,ds<\infty.
\]
The compensation formula therefore applies entrywise:
\[
 \begin{aligned}
 \E\int_0^tX_s\,d\widetilde C_{ab}(s)
 &=\E\int_0^tX_s\,dC_{ab}(s)
   -\E\int_0^tX_s\lambda_{ab}I_a(s-)\,ds\\
 &=0.
 \end{aligned}
\]
Consequently, by the definition of $N_i$,
\[
 \E\int_0^tX_s\,dN_i(s)
 =\sum_{j\ne i}\E\int_0^tX_s
 \bigl(d\widetilde C_{ji}(s)-d\widetilde C_{ij}(s)\bigr)=0.
\]

The drift terms in~\eqref{eq:product-integral}
are absolutely integrable by~\eqref{eq:local-moment}
and~\eqref{eq:cross-integrability}.
Taking expectations, applying Fubini's theorem,
and using the definitions in~\eqref{eq:moments},
we obtain
\begin{equation}\label{eq:moment-integral}
 \begin{aligned}
 M_i(t)=M_i(0)+\int_0^t\Bigl(
 &A_iM_i(s)+M_i(s)A_i^\top\\
 &+B_iU_i(s)+U_i(s)^\top B_i^\top\\
 &+\sum_j\lambda_{ji}M_j(s)+p_i(s)\Sigma
 \Bigr)\,ds.
 \end{aligned}
\end{equation}
Since each entry of the integrand
in~\eqref{eq:moment-integral} is integrable
on $[0,T]$, the fundamental theorem for
Lebesgue integrals implies that $M_i$
is absolutely continuous on $[0,T]$.
Differentiating~\eqref{eq:moment-integral}
yields~\eqref{eq:moment-equation}
for almost every $t\in[0,T]$.
Since $T>0$ is arbitrary, $M_i$ is locally
absolutely continuous.

Finally, the proof does not require the
positive definiteness of $\Sigma=DD^\top$.
Hence the conclusion remains valid
when $\Sigma$ is degenerate, including $D=0$.

\end{proof}

\section{Proof of Lemma~\ref{lem:conditional}}
\label{app:conditional}

\begin{proof}
Let $\Psi$ and $\Phi$ be as defined in the statement.
By the standard theory of fundamental matrices for
linear differential equations
\cite[Chapter~3]{CoddingtonLevinson1955},
applied pathwise to the bounded coefficient matrix
$F_{\theta_t,\ell}$, the matrix $\Psi(t)$ is invertible
for all $t\in[a,b]$.
Since $F_{\theta_t,\ell}$ is adapted,
both $\Psi$ and $\Psi^{-1}$ are adapted.
Moreover, the standard Gronwall estimates yield
\[
 \|\Phi(t,s)\|\le e^{M(t-s)},
 \qquad a\le s\le t\le b,
\]
and
\[
 \max\{\|\Psi(t)\|,\|\Psi(t)^{-1}\|\}
 \le e^{M(t-a)},\qquad t\in[a,b].
\]

Since $\Psi^{-1}$ is continuous and of finite variation,
the It\^o product formula
\cite[Section~4.3]{LiptserShiryayev1977} gives
\[
 d\bigl(\Psi(t)^{-1}z_t\bigr)
 =\Psi(t)^{-1}D\,dW_t.
\]
Integrating from $a$ to $t$ and multiplying by
$\Psi(t)$, we obtain
\[
 z_t=\Phi(t,a)z_a+\eta_t,
 \qquad
 \eta_t=\Psi(t)\int_a^t\Psi(s)^{-1}D\,dW_s.
\]
The stochastic integral is well defined in the
original filtration because its integrand is
adapted and uniformly bounded.

Fix $t\in[a,b]$ and let $\mathcal H$ be as in
the statement. Since $W$ and $\theta$ are
independent, the Brownian increments after time $a$
are independent of $\mathcal H$.

For a partition
$\Pi:\ a=s_0<\cdots<s_q=t$, define
\[
 \begin{aligned}
 \eta_t^\Pi
 &=\sum_{k=0}^{q-1}
   \Phi(t,s_k)D(W_{s_{k+1}}-W_{s_k})\\
 &=\Psi(t)\sum_{k=0}^{q-1}
   \Psi(s_k)^{-1}D(W_{s_{k+1}}-W_{s_k}).
 \end{aligned}
\]
The second expression is a bounded matrix
multiplied by an adapted It\^o sum.
Since $\Psi^{-1}$ is continuous and uniformly
bounded, the It\^o isometry
\cite[Section~4.2]{LiptserShiryayev1977}
and dominated convergence imply
\[
 \eta_t^\Pi\longrightarrow\eta_t
 \qquad\text{in }L^2
\]
as the mesh of $\Pi$ tends to zero.

Each $\Phi(t,s_k)$ is $\mathcal H$-measurable,
whereas the Brownian increments
$W_{s_{k+1}}-W_{s_k}$ are independent of
$\mathcal H$, have zero mean, and have covariance
$(s_{k+1}-s_k)I_r$.
Consequently,
\[
 \E[\eta_t^\Pi\mid\mathcal H]=0,
\]
and
\[
 \begin{aligned}
 \E\bigl[
   \eta_t^\Pi(\eta_t^\Pi)^\top
   \mid\mathcal H\bigr]=
 \sum_{k=0}^{q-1}
 \Phi(t,s_k)\Sigma\Phi(t,s_k)^\top
 (s_{k+1}-s_k).
 \end{aligned}
\]

The $L^2$ convergence of $\eta_t^\Pi$
implies $L^1$ convergence of the corresponding
matrix products. Furthermore, the continuity
and uniform boundedness of $\Phi(t,s)$ imply
that the Riemann sums on the right converge
in $L^1$ to
\[
 \int_a^t
 \Phi(t,s)\Sigma\Phi(t,s)^\top\,ds.
\]
Passing to the limit in the conditional
expectations therefore yields
\[
 \E[\eta_t\mid\mathcal H]=0,
 \qquad
 \E[\eta_t\eta_t^\top\mid\mathcal H]
 =\int_a^t
 \Phi(t,s)\Sigma\Phi(t,s)^\top\,ds.
\]
This completes the proof.
\end{proof}


\begin{thebibliography}{99}
\small
\setlength{\itemsep}{3pt}
\setlength{\parsep}{0pt}

\bibitem{Guo2020}
L.~Guo, ``Feedback and uncertainty: Some basic problems and results,''
\emph{Annual Reviews in Control}, vol.~49, pp.~27--36, 2020.
doi: \doi{10.1016/j.arcontrol.2020.04.001}.

\bibitem{SiqueiraTerra2009}
A.~A.~G. Siqueira and M.~H. Terra,
``A fault-tolerant manipulator robot based on
$H_2$, $H_\infty$, and mixed $H_2/H_\infty$
Markovian controls,''
\emph{IEEE/ASME Transactions on Mechatronics},
vol.~14, no.~2, pp.~257--263, 2009.
doi: \doi{10.1109/TMECH.2008.2009442}.

\bibitem{LiuMiaoFanHan2017}
Z.~Liu, S.~Miao, Z.~Fan, and J.~Han,
``Markovian switching model and non-linear
DC modulation control of AC/DC power system,''
\emph{IET Generation, Transmission \& Distribution},
vol.~11, no.~10, pp.~2654--2663, 2017.
doi: \doi{10.1049/iet-gtd.2016.1862}.

\bibitem{CostaFragosoTodorov2013}
O.~L.~V. Costa, M.~D. Fragoso, and M.~G. Todorov,
\emph{Continuous-Time Markov Jump Linear Systems}.
Berlin: Springer, 2013. doi: \doi{10.1007/978-3-642-34100-7}.

\bibitem{VargasCostaDoVal2013}
A.~N. Vargas, E.~F. Costa, and J.~B.~R. do Val,
``On the control of Markov jump linear systems with
no mode observation: Application to a DC motor device,''
\emph{International Journal of Robust and Nonlinear Control},
vol.~23, no.~10, pp.~1136--1150, 2013.
doi: \doi{10.1002/rnc.2911}.

\bibitem{Sworder1969}
D.~D. Sworder, ``Feedback control of a class of linear systems with jump parameters,''
\emph{IEEE Transactions on Automatic Control}, vol.~14, no.~1, pp.~9--14, 1969.
doi: \doi{10.1109/TAC.1969.1099088}.

\bibitem{JiChizeck1990}
Y.~Ji and H.~J. Chizeck,
``Controllability, stabilizability, and continuous-time Markovian jump linear quadratic control,''
\emph{IEEE Transactions on Automatic Control}, vol.~35, no.~7, pp.~777--788, 1990.
doi: \doi{10.1109/9.57016}.

\bibitem{FengEtAl1992}
X.~Feng, K.~A. Loparo, Y.~Ji, and H.~J. Chizeck,
``Stochastic stability properties of jump linear systems,''
\emph{IEEE Transactions on Automatic Control}, vol.~37, no.~1, pp.~38--53, 1992.
doi: \doi{10.1109/9.109637}.

\bibitem{FangLoparo2002}
Y.~Fang and K.~A. Loparo,
``Stabilization of continuous-time jump linear systems,''
\emph{IEEE Transactions on Automatic Control}, vol.~47, no.~10, pp.~1590--1603, 2002.
doi: \doi{10.1109/TAC.2002.803528}.

\bibitem{AitRamiElGhaoui1996}
M.~Ait Rami and L.~El Ghaoui,
``LMI optimization for nonstandard Riccati equations arising in stochastic control,''
\emph{IEEE Transactions on Automatic Control}, vol.~41, no.~11, pp.~1666--1671, 1996.
doi: \doi{10.1109/9.544005}.

\bibitem{FragosoCosta2005}
M.~D. Fragoso and O.~L.~V. Costa,
``A unified approach for stochastic and mean square stability of continuous-time linear systems with Markovian jumping parameters and additive disturbances,''
\emph{SIAM Journal on Control and Optimization}, vol.~44, no.~4, pp.~1165--1191, 2005.
doi: \doi{10.1137/S0363012903434753}.

\bibitem{CainesChen1985}
P.~E. Caines and H.-F. Chen,
``Optimal adaptive LQG control for systems with finite state process parameters,''
\emph{IEEE Transactions on Automatic Control}, vol.~30, no.~2, pp.~185--189, 1985.
doi: \doi{10.1109/TAC.1985.1103907}.

\bibitem{CainesNassiri1991}
P.~E. Caines and K.~Nassiri-Toussi,
``On the adaptive stabilization and ergodic behaviour of stochastic systems with jump-Markov parameters via nonlinear filtering,''
in \emph{Topics in Stochastic Systems: Modelling, Estimation and Adaptive Control},
Lecture Notes in Control and Information Sciences, vol.~161.
Berlin: Springer, 1991, pp.~185--215.
doi: \doi{10.1007/BFb0009305}.

\bibitem{DufourBertrand1994}
F.~Dufour and P.~Bertrand, ``Stabilizing control law for hybrid models,''
\emph{IEEE Transactions on Automatic Control}, vol.~39, no.~11, pp.~2354--2357, 1994.
doi: \doi{10.1109/9.333792}.

\bibitem{CainesZhang1995}
P.~E. Caines and J.-F. Zhang,
``On the adaptive control of jump parameter systems via nonlinear filtering,''
\emph{SIAM Journal on Control and Optimization}, vol.~33, no.~6, pp.~1758--1777, 1995.
doi: \doi{10.1137/S0363012992238679}.

\bibitem{DufourElliott1998}
F.~Dufour and R.~J. Elliott, ``Adaptive control of linear systems with Markov perturbations,''
\emph{IEEE Transactions on Automatic Control}, vol.~43, no.~3, pp.~351--372, 1998.
doi: \doi{10.1109/9.661591}.

\bibitem{HuangGuo2001}
M.~Huang and L.~Guo,
``Stabilization of stochastic systems with hidden Markovian jumps,''
\emph{Science in China Series F: Information Sciences}, vol.~44, no.~2, pp.~104--118, 2001.

\bibitem{XueGuoHuang2001}
F.~Xue, L.~Guo, and M.~Y. Huang,
``Towards understanding the capability of adaptation for time-varying systems,''
\emph{Automatica}, vol.~37, no.~10, pp.~1551--1560, 2001.
doi: \doi{10.1016/S0005-1098(01)00098-X}.

\bibitem{XueGuo2001}
F.~Xue and L.~Guo,
``Necessary and sufficient conditions for adaptive stabilizability of jump linear systems,''
\emph{Communications in Information and Systems}, vol.~1, no.~2, pp.~205--224, 2001.

\bibitem{LiuGuo2021}
N.~Liu and L.~Guo,
``A spectrum characterization of feedback capability for unknown Markovian jump linear systems,''
in \emph{Proceedings of the 40th Chinese Control Conference}, pp.~1050--1054, 2021.
doi: \doi{10.23919/CCC52363.2021.9550255}.

\bibitem{TanZhangYao2005}
S.~Tan, J.-F. Zhang, and L.~Yao,
``Optimality analysis of adaptive sampled control of hybrid systems with quadratic index,''
\emph{IEEE Transactions on Automatic Control}, vol.~50, no.~7, pp.~1044--1051, 2005.
doi: \doi{10.1109/TAC.2005.851456}.

\bibitem{TanZhang2009}
S.~Tan and J.-F. Zhang,
``Double-stepped adaptive control for hybrid systems with unknown Markov jumps and stochastic noises,''
\emph{ESAIM: Control, Optimisation and Calculus of Variations}, vol.~15, no.~4, pp.~969--993, 2009.
doi: \doi{10.1051/cocv:2008057}.

\bibitem{BercuDufourYin2009}
B.~Bercu, F.~Dufour, and G.~G. Yin,
``Almost sure stabilization for feedback controls of regime-switching linear systems with a hidden Markov chain,''
\emph{IEEE Transactions on Automatic Control}, vol.~54, no.~9, pp.~2114--2125, 2009.
doi: \doi{10.1109/TAC.2009.2026938}.

\bibitem{LiuLiuDong2010}
F.~Liu, Z.~Liu, and H.-B. Dong,
``Sufficient and necessary condition for stabilization of jump linear systems with noise,''
in \emph{Proceedings of the 2010 Chinese Control and Decision Conference},
pp.~2804--2807, 2010. doi: \doi{10.1109/CCDC.2010.5498704}.

\bibitem{LiuHeLi2016}
F.~Liu, H.~He, and P.~Li,
``Adaptive linear quadratic Gaussian control of Markov jump linear systems,''
in \emph{Proceedings of the 35th Chinese Control Conference},
pp.~1752--1757, 2016. doi: \doi{10.1109/ChiCC.2016.7553346}.

\bibitem{LiptserShiryayev1977}
R.~S. Liptser and A.~N. Shiryayev,
\emph{Statistics of Random Processes I: General Theory}.
New York: Springer-Verlag, 1977. doi: \doi{10.1007/978-1-4757-1665-8}.

\bibitem{MeynTweedie2009}
S.~Meyn and R.~L. Tweedie,
\emph{Markov Chains and Stochastic Stability}, 2nd~ed.
Cambridge: Cambridge University Press, 2009. doi: \doi{10.1017/CBO9780511626630}.

\bibitem{BaranYinZhu2013}
N.~A. Baran, G.~Yin, and C.~Zhu,
``Feynman-Kac formula for switching diffusions: connections of systems of partial differential equations and stochastic differential equations,''
\emph{Advances in Difference Equations}, vol.~2013, article~315, 2013.
doi: \doi{10.1186/1687-1847-2013-315}.


\bibitem{CoddingtonLevinson1955}
E.~A. Coddington and N.~Levinson,
\emph{Theory of Ordinary Differential Equations}.
New York: McGraw-Hill, 1955.

\end{thebibliography}
\end{document}